\documentclass[12pt]{article}
\usepackage{graphicx}
\usepackage[utf8]{inputenc}
\usepackage{amsmath}
\usepackage{amsfonts}
\usepackage{amssymb}
\usepackage{graphicx}

\usepackage[colorlinks=true, citecolor=magenta, linkcolor=blue,urlcolor=red]{hyperref}

\begin{document}

\begin{center}
\Large {\bf  A Note on Stress-Energy Tensor and Variational Principle for Null Strings}
\end{center}

\bigskip
\bigskip

\begin{center}
E.A. Davydov, D.V. Fursaev, V.A. Tainov
\end{center}

\bigskip
\bigskip

\begin{center}
{\it Dubna State University \\
     Universitetskaya st. 19\\
     141 980, Dubna, Moscow Region, Russia\\

  and\\

  the Bogoliubov Laboratory of Theoretical Physics\\
  Joint Institute for Nuclear Research\\
  Dubna, Russia\\}
 \medskip
\end{center}

\bigskip
\bigskip

\begin{abstract}
A straightforward application of the variational principle to null strings meets difficulties since string's world-sheets are degenerate. It is known that the variational principle in this case can be formulted with the help 
of two-vector density on the string world-sheet which plays a role of Lagrange multipliers.
It is shown that recently suggested stress-energy tensor of null strings can be derived 
by variation over the background metric of the action used to describe tensionless limit in the string theory.
One of the Lagrange multipliers is related to the energy of the null string.
\end{abstract}

\vspace*{20pt}

Null strings are hypothetical one-dimensional objects, whose points move orthogonally to the string along null geodesics \cite{Schild:1976vq}.
Null strings are interesting as quantum microscopic objects which describe a high energy or tensionless limit of the fundamental string theory \cite{Gross:1987kza}, \cite{Gross:1987ar}.
Null strings can be also considered as macroscopic cosmic strings which could have been formed in the early Universe. Such strings lead to a number of potentially observable physical effects in cosmology \cite{Fursaev:2017aap}, \cite{Fursaev:2018spa}. 

Since the world-sheet of a null string is degenerate its action cannot be defined in the Nambu-Goto form, as for tensile strings. 
Various definitions of the action for a null string have been proposed \cite{Zheltukhin:1989ar,Lindstrom:1993yb,Isberg:1993av} to describe the high energy limit of the string theory. 

Let $x^\mu = X^\mu(\lambda,\tau)$ be equations for the world-sheet of a null string, where $\lambda$ and $\tau$ are real parameters.
It allows a pair of tangent vectors, $l={dX / d\lambda}$, $\eta={dX / d\tau}$, $l$ is a 4-velocity of the string, $l^2=0$, and $\eta$ is a connecting vector, $\eta^2>0$.
The string moves in a space-time $\cal M$ with metric $g_{\mu\nu}$ in accord with the following equations \cite{Schild:1976vq}:
\begin{equation} \label{i.2}
(l \cdot \eta)=0 ~~,
\end{equation}
\begin{equation} \label{i.3}
\nabla_l l = \beta l~~. 
\end{equation}
Consider a string-like object on $\cal M$ with the trajectory $x^\mu = X^\mu(\zeta)$ parametrized by a couple of 
real parameters $\zeta^a$, $a=1,2$. A priory one does not require that $X^\mu(\zeta)$ describe a null string. Instead, it is expected that
(\ref{i.2}), (\ref{i.3}) along with condition $l^2=0$ appear as stationary points of an action.
One of the options for such a string action is \cite{Lindstrom:1993yb}
\begin{equation}\label{1}
I[X,V,g] =\frac{1}{2}\int d^2\zeta~V^\mu(x) g_{\mu\nu}V^\nu(x)~~,
\end{equation}
where $V^\mu = V^a {X^\mu_{,a}}$, $X^\mu_{,a}={dX^\mu / d\zeta^a}$. The additional dynamical variable $V^a=V^a(\zeta)$ is a world-sheet 2-vector density 
which plays a role of Lagrange multipliers. The nice property of (\ref{1}) is that it is coordinate and reparametrization invariant under proper transformations of $V$. The string action is required, for example, in quantum theory of null strings.

In the cosmological context, when null strings are considered as macroscopic objects, one needs the stress-energy tensor (SET) of strings, 
since it is a source of its gravitational field. The SET of null cosmic strings has been recently proposed in \cite{Davydov:2022qil}
in the form:
\begin{equation}\label{set}
T^{\mu\nu}(x) = \int d\lambda d\tau \bar{\mu}(\lambda,\tau) l^\mu l^\nu \frac{\delta^{(4)}(x-X(\lambda,\tau))}{\sqrt{-g}}~~.
\end{equation}
The delta-function indicates that $T^{\mu\nu}(x)$  is localized on the world-sheet of the string. The quantity $\bar{\mu}d\tau$ is 
energy of a string segment between $\tau$ and $\tau+d\tau$. Definition (\ref{set}) has the following properties:
i)  each string segment contributes to SET as a massless particle;
ii) $T_\mu^\mu=0$; iii) the covariant conservation law $\nabla_\mu T^{\mu\nu}=0$ holds under the
condition
\begin{equation}\label{en}
\partial_\lambda \bar{\mu} + \beta \bar{\mu}=0~~,
\end{equation}
where $\beta$ is defined in (\ref{i.3}). The statements of this note are the following:
\begin{enumerate}
	\item string equations are stationary points of (\ref{1}),  that is, variations $\delta_XI[X,V,g]=\\ \delta_VI[X,V,g]=0$ under fixed background metric
	$g$ yield $l^2=0$ and (\ref{i.2}), (\ref{i.3});
	\item stress-energy tensor (\ref{set}) can be derived from (\ref{1}) by metric variation 
	\begin{equation}\label{set2}
		T^{\mu\nu}(x) ={2 \over \sqrt{-g}}{\delta I[X,V,g] \over \delta g_{\mu\nu}(x)}~~,
	\end{equation}
	where after variation is performed $X$ and $V$ in (\ref{set2}) are taken on-shell. 
\end{enumerate}
To prove the above statements it is convenient to take components of the world-sheet vector as
$V^a = \sqrt{\bar{\mu}_1}(1,\bar{\mu}_2)$, where  $\bar{\mu}_k = \bar{\mu}_k(\zeta)$.
One can write (\ref{1}) in the form:
\begin{equation} \label{1.2}
I[X,V(\bar{\mu}),g] = \frac{1}{2} \int d^2\zeta~ \bar{\mu}_1 (v\cdot v)~~,
\end{equation}
where $v=e_1+\bar{\mu}_2 e_2$ and $e_k^\mu=dX^\mu / d\zeta^k$.
If the background metric is fixed,
stationary configurations of (\ref{1.2}) are determined by equations
\begin{equation} \label{1.3a}
(v\cdot v)=0~~,
\end{equation}
\begin{equation} \label{1.3b}
(e_2\cdot v)=0~~,
\end{equation}
\begin{equation}\label{1.3c}
\nabla_v v=\bar{\beta}v~~,~~\bar{\beta}=-\bar{\mu}_1^{-1}(\bar{\mu}_{1,\zeta^1}+(\bar{\mu}_1\bar{\mu}_2)_{,\zeta^2})~~.
\end{equation}
Eqs. (\ref{1.3a}), (\ref{1.3b}) are determined by variations over $\bar{\mu}_k$, and (\ref{1.3c}) over $X$.
Application of (\ref{set2}) to \eqref{1.2} yields the stress-energy tensor
\begin{equation} \label{1.3}
T^{\mu\nu} (x)= \int d^2\zeta~ 
\bar{\mu}_1 v^\mu v^\nu
\frac{\delta^{(4)}(x-X(\zeta))}{\sqrt{-g}}~~.
\end{equation}
It follows from (\ref{1.3a}), (\ref{1.3c}) that vector $V^\mu$ is a null tangent vector to the world-sheet which generates null geodesics.
Therefore the world-sheet is degenerate, and $V$ can be interpreted as the velocity of the string segment.
Vector $e_2$ can be interpreted as the connecting vector.
Equation (\ref{1.3b}) fixes one of the components of the 2-vector $v^a$, $\bar{\mu}_2 = - \left(e_1 \cdot e_2\right)/e_2^2$, 
while the other component $\bar{\mu}_1$ turns out to be related to the string energy in (\ref{1.3}).

Since  $V^a$ is a 2-vector density, one can chose the "gauge" $V^{a=2}=\bar{\mu}_2=0$. This implies that 
$v^\mu=l^\mu$, $e_2^\mu=\eta^\mu$, $\zeta^1=\lambda$, $\zeta^2=\tau$. The second equation in (\ref{1.3c})  also implies (\ref{en}) where $\beta=\bar{\beta}$ and $\bar{\mu}=\bar{\mu}_1$. The above gauge
condition can be realized when $\zeta^1$ is chosen to parametrize points along null geodesics 
on the string world-sheet generated by $v^\mu$.

The fact that auxiliary vector $V$, which has been introduced in \cite{Lindstrom:1993yb} to formulate variational principle
for equations of null strings, is related to string energy is quite interesting.
The physical energy $\mu$ of the string per unit length is defined as $\mu = \bar{\mu} |\eta|^{-1}$, see \cite{Davydov:2022qil},
and it does not depend on reparametrizations of $\tau$. The conservation law for physical energy is
\begin{equation} \label{1.9}
\partial_\lambda \mu + \theta \mu =0~~,
\end{equation}
where $\theta = \partial_\lambda \ln |\eta|$ is expansion parameter of the string world-sheet. A distinctive feature of null cosmic strings
 is in their optical properties. In addition to the expansion parameter one can introduce the rotation parameter $\kappa$ and complement
(\ref{1.9}) with an analogue of the  Sachs optical equations \cite{Fursaev:2021xlm}:
\begin{equation}\label{i.11}
\partial_\lambda Z+Z^2=-\Psi_0-\Phi_{00}~~,
\end{equation}
where $Z=\theta+i\kappa$ is a spin coefficient, $\lambda$ is the affine parameter on the string trajectory, $\Psi_0$, $\Phi_{00}$ are invariants constructed from the components of the Weyl and Ricci tensors, respectively.

The above analysis implies that the string does not have caustics where $\eta^2=0$.  

For future applications, it seems important that the variational principle introduced to describe tensionless limit of 
the string theory \cite{Lindstrom:1993yb} can be extended to derive the metric stress-energy tensor of null strings which can be used to study physical effects of the cosmic strings in cosmology. It is important that the string energy per unit length $\bar{\mu}$ is determined by 2-vector density $V$, which has been considered so far as a purely auxiliary variable.

\section*{Acknowledgments}

This research is supported by Russian Science Foundation grant No. 22-22-00684, \\
https://rscf.ru/project/22-22-00684/.


\bigskip
\bigskip
\bigskip



\newpage


\begin{thebibliography}{}

\bibitem{Schild:1976vq}
A.~Schild,
``Classical Null Strings,''
Phys. Rev. D \textbf{16}, 1722 (1977)
doi:10.1103/PhysRevD.16.1722

\bibitem{Gross:1987kza}
D.~J.~Gross and P.~F.~Mende,
``The High-Energy Behavior of String Scattering Amplitudes,''
Phys. Lett. B \textbf{197}, 129-134 (1987)
doi:10.1016/0370-2693(87)90355-8

\bibitem{Gross:1987ar}
D.~J.~Gross and P.~F.~Mende,
``String Theory Beyond the Planck Scale,''
Nucl. Phys. B \textbf{303}, 407-454 (1988)
doi:10.1016/0550-3213(88)90390-2


\bibitem{Fursaev:2017aap}
D.~V.~Fursaev,
``Physical effects of massless cosmic strings,''
Phys. Rev. D \textbf{96}, no.10, 104005 (2017)
doi:10.1103/PhysRevD.96.104005
[arXiv:1707.02438 [gr-qc]].



\bibitem{Fursaev:2018spa}
D.~V.~Fursaev,
``Massless Cosmic Strings in Expanding Universe,''
Phys. Rev. D \textbf{98}, no.12, 123531 (2018)
doi:10.1103/PhysRevD.98.123531
[arXiv:1811.01563 [gr-qc]].

\bibitem{Zheltukhin:1989ar}
A.~A.~Zheltukhin,
``Interaction of Null Strings and Null Membranes With Antisymmetric Tensor Fields,''
Phys. Lett. B \textbf{233}, 112-116 (1989)
doi:10.1016/0370-2693(89)90625-4


\bibitem{Lindstrom:1993yb}
U.~Lindstrom,
``The Zero tension limit of strings and superstrings,''
[arXiv:hep-th/9303173 [hep-th]].

\bibitem{Isberg:1993av}
J.~Isberg, U.~Lindstrom, B.~Sundborg and G.~Theodoridis,
``Classical and quantized tensionless strings,''
Nucl. Phys. B \textbf{411}, 122-156 (1994)
doi:10.1016/0550-3213(94)90056-6
[arXiv:hep-th/9307108 [hep-th]].

\bibitem{Davydov:2022qil}
E.~A.~Davydov, D.~V.~Fursaev and V.~A.~Tainov,
``Null cosmic strings: Scattering by black holes, optics, and spacetime content,''
Phys. Rev. D \textbf{105}, no.8, 083510 (2022)
doi:10.1103/PhysRevD.105.083510
[arXiv:2203.02673 [gr-qc]].

\bibitem{Fursaev:2021xlm}
D.~Fursaev,
``Optical equations for null strings,''
Phys. Rev. D \textbf{103}, no.12, 123526 (2021)
doi:10.1103/PhysRevD.103.123526
[arXiv:2104.04982 [gr-qc]].




\end{thebibliography}
\end{document}